\documentclass[usenatbib]{article}  

\usepackage{newtxtext,newtxmath}
\usepackage[T1]{fontenc}
\usepackage{graphicx}
\usepackage{subcaption} 

\title{The Third Generation of Nanogenerators: The Irreplaceable Potential Source Enabled by the Flexoelectric Nanogenerator}

\author{Shang Ru Li\thanks{These authors contributed equally.} \\ Qi Kang Zhang\thanks{These authors contributed equally.} \\ Xiao Xiong Wang\thanks{Corresponding author: wangxiaoxiong@qdu.edu.cn}}

\date{School of Physics, Qingdao University, China}  
\newenvironment{keywords}{\par\noindent\textbf{Keywords:}}{\par}

\begin{document}
\maketitle

\begin{abstract}
The electroneutrality assumption has long been adopted by scholars; however, this assumption may lead to an oversight of certain physical effects. Using derivations from a discontinuous medium, we have obtained an expression for the potential and energy of a many-body unipolar charge system, which corresponds well to its counterpart in a continuous medium. The compressed form of this expression suggests that compressing a macroscale charged body to the nanoscale can yield an enormous electric potential and energy, thereby establishing a concrete research framework for third-generation nanogenerators. This effect may serve as a crucial reference for understanding anomalous spatial electromagnetic distributions and divergent energy fields.
\end{abstract}




\section{Introduction}

Research in electricity has long been based on the assumption of electroneutrality \cite{MacGillivray1968, Smith1993}. For instance, in calculations involving parallel plate capacitors \cite{Nishiyama1994, Grove2005}, it is implicitly assumed that the charge on the grounded electrode exactly counterbalances that on the charged electrode. However, non-electroneutral effects are far from rare. In recent years, with the rapid development of nanogenerators, numerous studies have begun to document interactions among highly charged systems. Drawing on our extensive measurement experience \cite{He2019, Yang2019}, we observe that the strength of electrostatic induction decays with distance, indicating that the induced charge on an electrode is not merely the opposite of the charge on the charged body, as is often simplified in textbooks.

We posit that this phenomenon deviates from the mainstream academic consensus, which asserts that "the universe has undergone sufficient relaxation to achieve overall electroneutrality, and any minor non-neutral charge distributions occur only locally." Instead, we argue that the widely accepted assumption of electroneutrality primarily arises from limitations in measurement techniques. For example, a traditional voltmeter essentially amplifies the current passing through a measurement circuit rather than integrating the spatial electrostatic flux, rendering it incapable of detecting static charge distributions. A hallmark characteristic of non-electroneutral effects is the presence of a net electrostatic flux. By contrast, under electroneutral conditions—such as those in a dipole moment—the net flux at an infinite distance is conventionally negligible. As an illustrative example, the Earth carries approximately \(5 \times 10^5\) C of charge \cite{Rycroft2012, Harrison2004}; if only the static electric field at a person’s height is considered, it could result in an effective potential difference of approximately 350 V from foot to head. While this potential remains unobservable in a static environment, it can be detected in a rapidly rotating electrostatic induction system. Mechanisms such as contact electrification and thermionic emission are among the primary contributors to non-electroneutral phenomena. Despite substantial theoretical and experimental evidence, many researchers continue to assume electroneutrality by default, without further scrutiny.

Although the governing laws of non-electroneutral systems \cite{Morf2010,Lyahov2014} can be logically constructed, centuries of research have yet to yield fully satisfactory dielectric constraints, and experiments involving static charges frequently encounter stability challenges. Consequently, identifying an effective physical mechanism as an entry point is crucial, as it provides a viable experimental pathway for verification. Moreover, classical electromagnetism textbooks frequently present methods for calculating the potential of continuously charged bodies, such as the inverse relationship between the potential of a charged sphere and its radius \cite{Waisman1972, Wu2019}. However, ensuring the consistency of this compression effect in both continuous and discontinuous media is essential for determining the system state under practical ionic-carrier conditions, as well as for investigating the potential onset of field emission.

In this work, we begin by addressing the divergence of the Madelung constant \cite{Borwein1985,Crandall1987} in discrete media to establish the extensive property of potential in a unipolar charge system. We reinterpret current non-electroneutral phenomena as contributions to potential energy within the framework of free energy discussions and derive a corresponding analytical expression, which we validate through experimental fitting. Extrapolation of this expression suggests that compressing a macroscale charged body to the nanoscale—such as in a dilute plasma—can yield gigavolt-level potentials and exceptionally high energy densities. Given the energy densities observed in first-generation piezoelectric and second-generation triboelectric nanogenerators, we contend that our findings constitute a landmark advancement in the design of third-generation nanogenerators \cite{Lyahov2014, Wu2019,Jiang2013}, while also providing crucial insights into anomalously high energy densities observed in space.

\section{Results and Discussion}

\subsection{Madelung Constant}

In many-body interactions, an important approach is to extend two-body interactions using the Madelung constant \cite{Argyriou1992}. The Madelung constant is a dimensionless parameter that depends solely on the crystal structure and essentially represents the bonding energy density of an ideal crystal. For commonly encountered crystals, the Madelung constant is relatively easy to calculate, and current research focuses on improving computational speed and deriving analytical expressions.

Because cations and anions in a crystal are arranged in a periodically alternating pattern, the Madelung constant tends to converge. However, in the case of like charges, whether in a one-dimensional ionic chain or a two-dimensional arrangement, the Madelung constant tends to diverge. 

Taking the NaCl structure as an example, we place a positive ion at the center, with \( r \) denoting the distance between adjacent ions. and then determining the interaction energy between this ion and all other surrounding ions accordingly.
\begin{equation}
    W = \frac{1}{4\pi\epsilon_0} \sum_{n_1 n_2 n_3} \frac{e^2 (-1)^{n_1+n_2+n_3}}{(n_1^2 r^2 + n_2^2 r^2 + n_3^2 r^2)^{1/2}}
\end{equation}

The Madelung constant is expressed as

\begin{equation}
    M = \sum_{n_1 n_2 n_3} \frac{(-1)^{n_1+n_2+n_3}}{(n_1^2 + n_2^2 + n_3^2)^{1/2}}
\end{equation}

\begin{figure}[htbp]
    \centering
    \begin{subfigure}[b]{0.48\textwidth}
        \includegraphics[width=\textwidth]{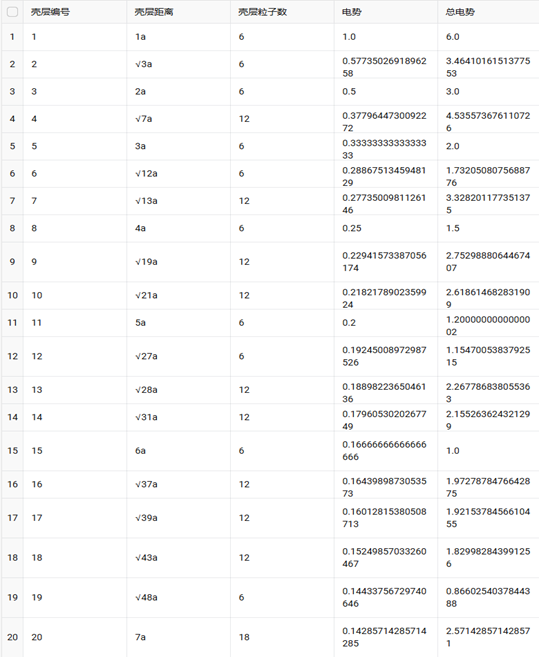}
        \caption{Parameters for each shell in the hexagonal structure.}
        \label{fig:shell_parameters}
    \end{subfigure}
    \hfill
    \begin{subfigure}[b]{0.48\textwidth}
        \includegraphics[width=\textwidth]{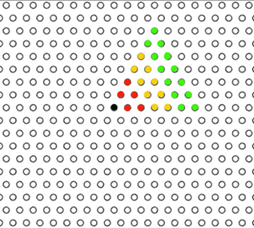}
        \caption{Visualization of charge distribution.}
        \label{fig:hexagonal_distribution}
    \end{subfigure}
    \caption{Parameters for each shell in the hexagonal close-packed structure and corresponding visualization.}
    \label{fig:hexagonal_figure}
\end{figure}

In the NaCl crystal structure, cations and anions alternate, resulting in a convergent electrostatic sum. However, if we consider an arrangement of ions with the same charge, the cumulative potential diverges. 

To illustrate this, we examine a two-dimensional hexagonal close-packed structure, where the pairwise electrostatic potential is inversely proportional to the distance between charges, and the distance between the central charge and each nearest neighbor is set to 1. Consequently, the potential contribution of a single charge at distance 1 from the central charge is 1. 

We begin with a central charge and divide the surrounding charges into tiers according to their distance from the center. For simplicity, the first through third nearest shells, containing a total of 18 charges (shown in red), are grouped into the first tier, which can be visualized as the two smallest concentric hexagons around the center. The second tier encompasses the fourth through eighth shells, containing 42 charges (shown in yellow), corresponding to the next two concentric hexagons. The third tier includes the ninth through fifteenth shells, containing 66 charges (shown in green), and so forth.

To compute the total potential of each tier, we note that the contribution of a single charge is inversely proportional to its distance from the center, and thus the total potential of a shell is obtained by multiplying the single-charge contribution by the number of charges in that shell. Mathematically, for a given tier \( n \):

\begin{equation}
    U_n = \sum_{i=1}^{N_n} \frac{1}{r_i}
\end{equation}

where \( N_n \) is the total number of charges in the \( n \)-th shell, and \( r_i \) represents the distance of each charge from the center.

From the collected data, the first tier contributes a total potential of 12.4, the second tier contributes 13.0, and the third tier contributes 13.1, indicating a slowly increasing trend. 

To verify whether this sum diverges, we apply a term-limit test: for the \( n \)-th tier, the total number of charges is:

\begin{equation}
    N_n = (4n - 1) \times 6
\end{equation}

The farthest distance of these charges from the center is \( 2n \), and thus each charge contributes at least:

\begin{equation}
    \frac{1}{2n}
\end{equation}

Consequently, multiplying by the total number of charges in that tier gives:

\begin{equation}
    \left(4n - 1\right) \times 6 \times \frac{1}{2n}
\end{equation}

which approaches 12 as \( n \to \infty \). Since this contribution remains finite and does not diminish to zero, the overall sum of the potential diverges when summed over all tiers.

\begin{figure}[htbp]
    \centering
    \includegraphics[width=0.8\textwidth]{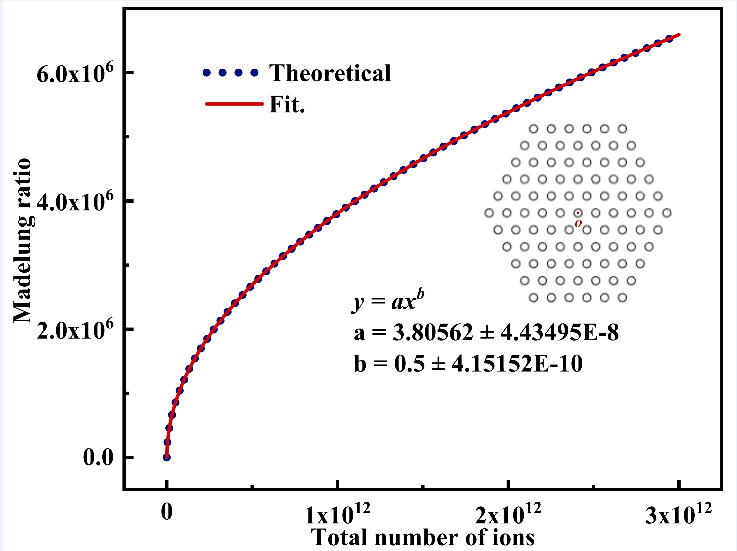} 
    \caption{Theoretically calculated and fitted results of the Madelung constant for a single-charge structure in a planar hexagonal close-packed arrangement.}
    \label{fig:madelung}
\end{figure}

The calculated values of the Madelung constant are shown in Figure~\ref{fig:madelung}.

As the total number of single-type ions increases, the Madelung constant also increases continuously. Through curve fitting, we obtained an expression for the non-convergent Madelung coefficient as a function of the total number of ions, given by:

\begin{equation}
    M = 3.80562 N^{0.5}
\end{equation}

In practical scenarios, the total number of ions \( N \) is determined by the ratio of the total charge \( Q \) to the elementary charge \( e \), i.e.,

\begin{equation}
    N = \frac{Q}{e}
\end{equation}

allowing the corresponding Madelung coefficient to be calculated. Consequently, the interaction energy between the central ion and all other surrounding ions can be expressed as:

\begin{equation}
    w = \frac{1}{4\pi\varepsilon_0} \frac{e^2}{r} M(N)
\end{equation}

This equation shows that the energy is inversely proportional to the interionic distance, implying that compressing a charged body will increase the system's energy.

\subsection{Compression of a Charged Body}

\textbf{Special Case:} For a uniformly charged sphere with a radius of \( R \) and a total charge of \( +Q \), the spatial electric field distribution can be obtained based on Gauss's theorem.

\begin{equation}
    E =
    \begin{cases} 
        \frac{1}{4\pi\varepsilon_0} \frac{Q}{r^2}, & \quad (r > R) \\[10pt]
        \frac{1}{4\pi\varepsilon_0} \frac{Q r}{R^3}, & \quad (r < R)
    \end{cases}
\end{equation}

By integration, the electric potential \( U_O \) at the center of the sphere (O point) is obtained.

\begin{equation}
    U_O = \frac{3}{2} \frac{Q}{4\pi\varepsilon_0} R^{-1}
\end{equation}

It can be seen that when the charge \( Q \) of the sphere remains constant, compressing the sphere by reducing its radius \( R \) results in an increase in the electric potential at the center as \( R \) decreases. If the compressed radius \( R_2 \) is reduced to half of the initial radius \( R_1 \), the electric potential at the center \( U_2 \) doubles compared to the initial potential \( U_1 \).

We performed finite element simulations to calculate the variations in the electric field and potential distribution from the center of the sphere to distant space before and after compression, as shown in figure3. The total charge remains unchanged before and after compression. In the region where \( r > R_1 \), both the electric field and potential remain unaffected, while in the region where \( r < R_1 \), the electric field becomes stronger, and the potential increases significantly compared to the pre-compression state.

During compression, as the sphere’s radius continuously decreases (from \( R = 10 \) m to \( R = 1 \) m), the finite element simulation results of the potential variation at point \( O \) are shown in figure3. By fitting the results to a formula, it is observed that the potential follows an inverse proportional relationship with the radius:

\begin{equation}
    U \propto R^b, \quad \text{where } b \approx -1.
\end{equation}

The fitting coefficient \( a \) is correlated with the charge \( Q \). In the simulation, we set \( Q = 1 \) pC, and the fitting result of coefficient \( a \) matches well with the theoretical value (\(\sim 0.01348\)) calculated from Equation11.

\begin{figure}[htbp]
    \centering
    \includegraphics[width=0.9\textwidth]{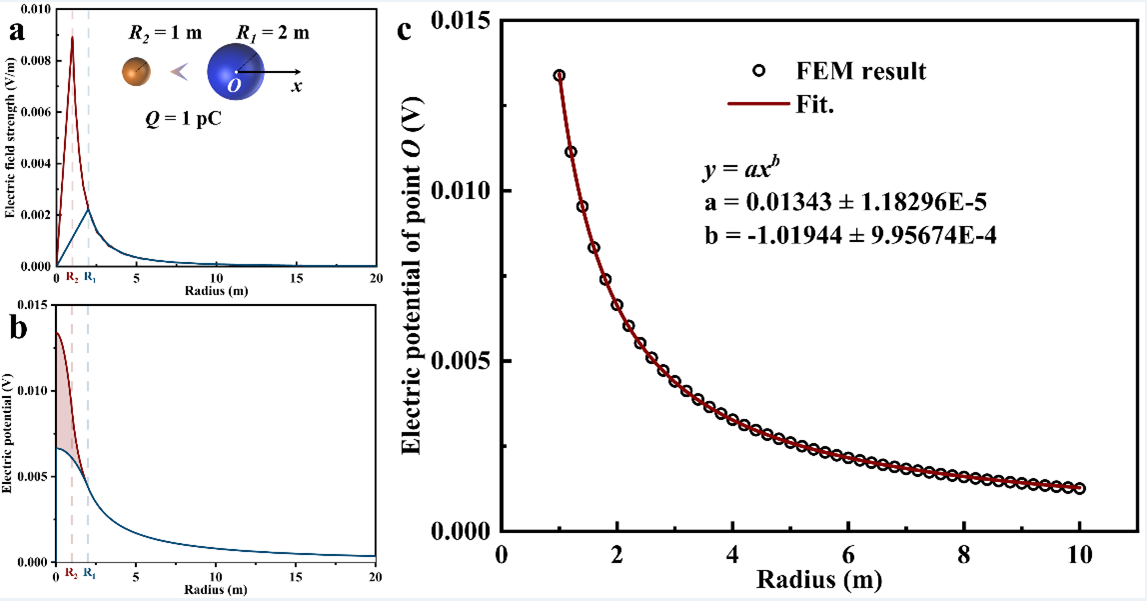}
    \caption{Finite element simulation results of the electric potential and field distribution changes before and after the compression of a uniformly charged sphere. 
    (a) Electric field distribution from the sphere’s center (point O) to distant space. 
    (b) Electric potential distribution from point O to distant space. 
    (c) Variation of the electric potential at point O with the compression of the sphere's radius and the corresponding fitting.}
    \label{fig:finite_element_simulation}
\end{figure}

For the general case, deriving an analytical expression is overly complex. Instead, we use finite element simulations to determine the approximate variation trend. As shown in figure4, we simulate the potential change at point \( O \) during the compression of two-dimensional square and circular charged bodies. As the side length or radius decreases, the potential at point \( O \) exhibits an inverse proportional increase, similar to the case of the sphere discussed earlier.

\begin{figure}[htbp]
    \centering
    \includegraphics[width=0.9\textwidth]{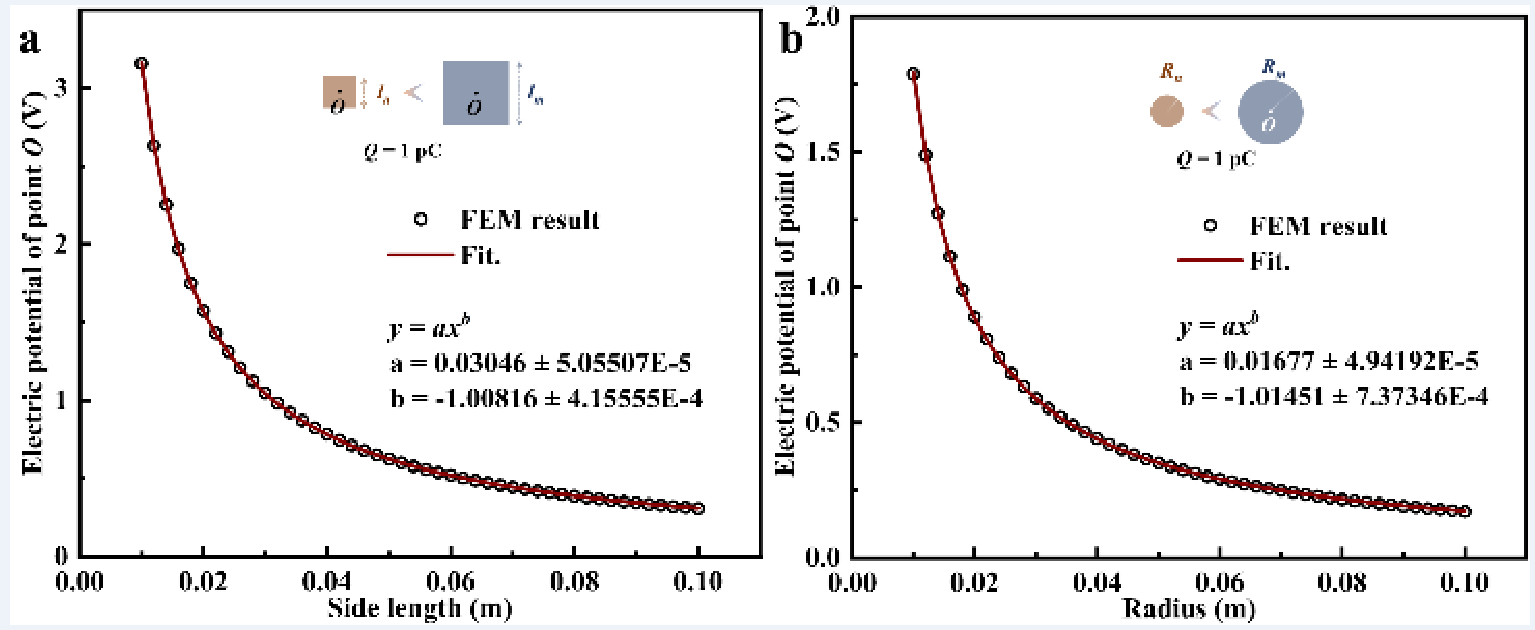}
    \caption{Variation of the electric potential at point \( O \) during the compression of a two-dimensional charged body. 
    (a) Electric potential at point \( O \) as a function of side length for a two-dimensional square, along with the fitting result. 
    (b) Electric potential at point \( O \) as a function of radius for a two-dimensional circle, along with the fitting result.}
    \label{fig:figure4}
\end{figure}

We conducted a quantitative experiment on compressed potential. While the electric potential at the center of a uniformly charged sphere can be easily expressed analytically, its direct measurement is challenging. Therefore, we performed a stretching and compression potential experiment using a strip-shaped charged body made of naturally stretchable latex. 

The experiment employed a non-contact electrostatic field meter to measure the potential at the center in real time. First, the strip was stretched to approximately 40 cm in length and charged by repeatedly making contact with and separating it from a PTFE film, ensuring that its surface acquired a positive charge. The electrostatic field meter was then placed 2.5 cm away from point \( O \) to measure the potential at that point. 

During the experiment, the strip was gradually compressed toward its center while maintaining the relative position between point \( O \) and the electrostatic field meter. The real-time variations in \( O \)'s potential were continuously displayed and recorded by the electrostatic field meter. From this experiment, we obtained the relationship between \( O \)'s potential and the strip length, as shown in figure5. The results indicate that as the length of the strip-shaped charged body decreases, the potential at point \( O \) follows an inverse proportional increase.

Based on the relationship between the potential and the radius of a charged sphere, we can calculate scenarios that are currently difficult to measure experimentally. For example, if the initial radius of a charged sphere is \( r_0 = 1 \) mm and the initial potential is \( U_0 = 10 \) kV, then the coefficient can be determined as:

\begin{equation}
    a = \frac{U_0}{r_0^{-1}} = 10.
\end{equation}

When the charged sphere is compressed to \( r_1 = 100 \) nm, the potential increases to:

\begin{equation}
    U_1 = a \cdot r_1^{-1} = 10^8 \text{ V}.
\end{equation}

Using the capacitor energy storage formula:

\begin{equation}
    W = \frac{Q^2}{2C} = \frac{1}{2} C U^2,
\end{equation}

we can estimate that after compression, the energy density of the system increases to the order of \( 10^{10} \) GJ/m\(^3\).

\begin{figure}[htbp]
    \centering
    \includegraphics[width=0.8\textwidth]{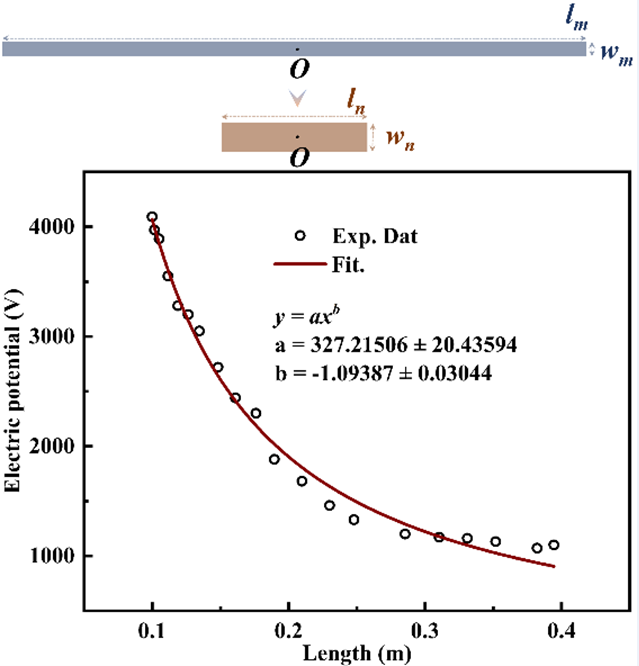}
    \caption{Experimental and fitted results of the variation in the potential at point \( O \) with the length of the strip-shaped charged body during stretching and compression.}
    \label{fig:figure5}
\end{figure}

\section{SUMMARY and OUTLOOK}

This work is based on the non-continuous dielectric assumption, that is, the non-Landau assumption. In a dielectric system where ions (plasma) serve as the basic unit, the following findings are made:
(a) The non-convergent Madelung constant is transformed into a Madelung coefficient, so the potential becomes an extensive total potential that depends on volume: 
\[
\varphi_{\text{total}} = \varphi_{\text{point}} \times \text{Madelung coefficient}.
\]
(b) The formula for the Madelung coefficient contains a distance-dependent term, leading to a higher potential when compression occurs. Due to the \( 1/r \) relationship between charge density and point potential, this compressed potential diverges as the volume approaches zero, effectively reflecting an increase in potential caused by the rising charge density.
(c) As \( r \) approaches zero, both the potential and charge density diverge, resulting in infinite energy density. This implies the potential to generate enormous energy, suggesting that this effect could serve as an energy source.
(d) Taking an initial potential of 10 kV with an initial volume of 1 mm³, the energy density obtained when the system is compressed to a volume of 100 nm³ is \( 10^{10} \) GJ/m³, which exceeds the energy density of all current human energy sources. Therefore, this can be labeled as a "third-generation nanogenerator". This massive energy source could potentially explain the extremely high temperature of the Sun's surface and the enormous energy field required to form magnetic loops.

Furthermore, in Figure 2, the red region represents the spatial potential change during compression, corresponding to the electric field variations between the two curves shown in Figure 2. These field changes provide dynamic boundary conditions; when the spatial period is shorter than the electromagnetic wavelength, they can disrupt decoherence symmetry and yield effective electromagnetic emission. However, this topic is not the central focus of the present work.

\bibliographystyle{unsrt}  
\bibliography{references}  



\label{lastpage}
\end{document}